\documentclass[a4paper]{spie}  
\usepackage[]{graphicx}

\title{The Physical Model in Action: Quality Control for X-Shooter} 

\author{Sabine Moehler\supit{a}, Paul Bristow\supit{a}, Florian Kerber\supit{a}, Andrea Modigliani\supit{a}, Joel Vernet\supit{a}
\skiplinehalf
\supit{a}European Southern Observatory, Karl-Schwarzschild-Str. 2,
85748 Garching, Germany \\
}

\begin{document} 
  \maketitle 

\begin{abstract}

The data reduction pipeline for the VLT 2$^{\rm nd}$
generation instrument X-Shooter uses a physical model to determine the
optical distortion and derive the wavelength calibration. The
parameters of this model describe the positions, orientations, and
other physical properties of the optical components in the
spectrograph. They are updated by an optimisation process that ensures
the best possible fit to arc lamp line positions. ESO Quality Control
monitors these parameters along with all of the usual
diagnostics. This enables us to look for correlations between inferred
physical changes in the instrument and, for example, instrument
temperature sensor readings.
\end{abstract}


\keywords{quality control, X-Shooter, ESO, VLT}

\section{INTRODUCING X-SHOOTER}
\label{sec:intro}  

\begin{figure}
\begin{center}
\begin{tabular}{c}
\includegraphics[width=0.7\textwidth]{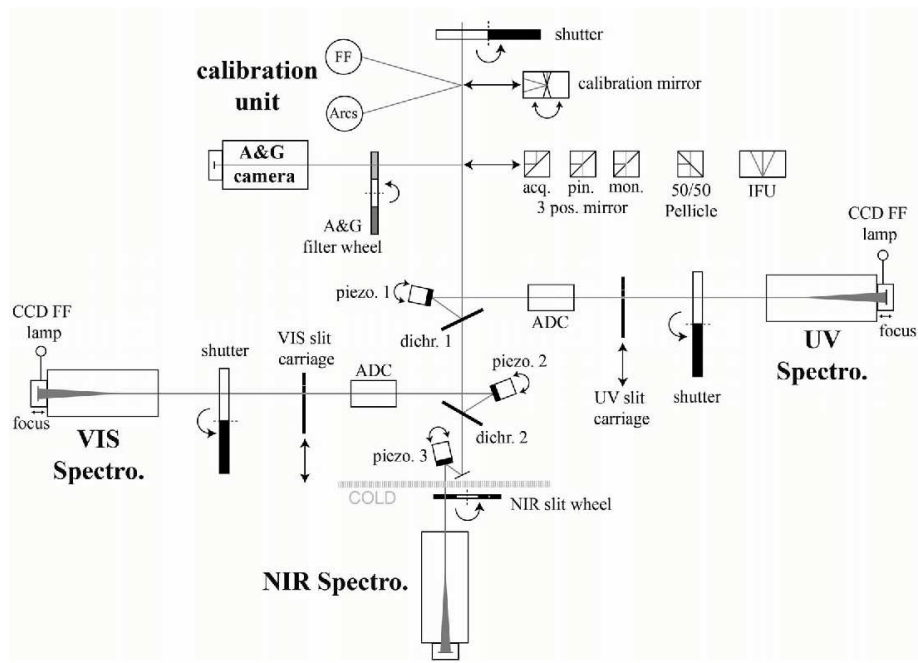}
\end{tabular}
\end{center}
\caption[]{Schematic layout of the X-Shooter instrument. The physical
  model includes the optical components after the ADC for the UVB and
  VIS arms and within the cold part of the NIR arm.\label{Fig:layout} }
\end{figure}

\begin{figure}
\begin{center}
\begin{tabular}{c}
\includegraphics[width=\textwidth]{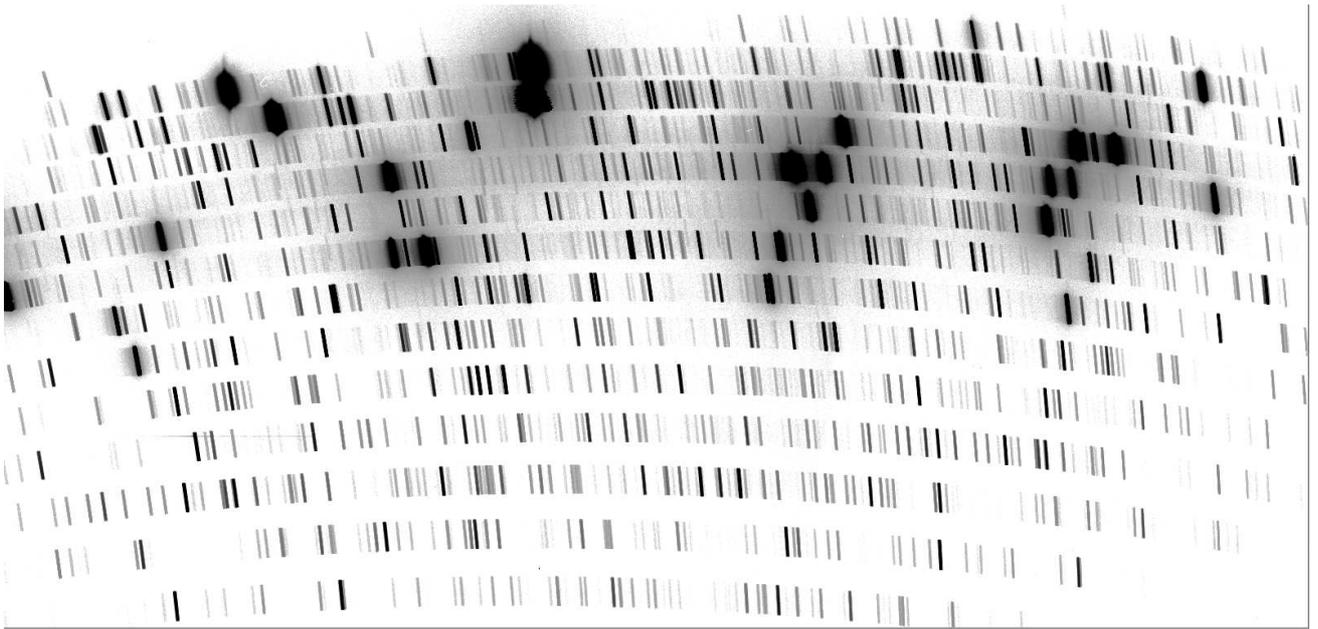}
\end{tabular}
\end{center}
\caption[]{VIS arc line frame. This image illustrates the order
curvature and line tilt present in X-Shooter data.
\label{Fig:tilt} }
\end{figure}

X-Shooter\cite{XSH_PROC} is the first of the second generation
instruments at the ESO Very Large Telescope (VLT) and has been
available since October 1, 2009. It is a medium resolution
spectrograph covering the wavelength range of 300--2500\,nm in three
arms (see Fig~\ref{Fig:layout} for a schematic layout). This allows
each arm to be optimised for the wavelength range it covers. The arms
are UVB (300--559\,nm, $\lambda/\Delta \lambda$ = 5100 for a
1$\stackrel{''}{}$ slit), VIS (550--1020\,nm, $\lambda/\Delta \lambda$
= 7700 for a 1$\stackrel{''}{}$ slit), and NIR (1020--2480\,nm,
$\lambda/\Delta \lambda$ = 4800 for a 1$\stackrel{''}{}$ slit). The
separation of light between the arms is achieved with two
dichroics. In order to combine large wavelength coverage with medium
resolution cross-dispersed echelle spectrographs are used. This
results in strongly curved orders with highly tilted lines in each
order. The line tilt varies with line position along the order (see
Fig.~\ref{Fig:tilt} for an example in the VIS arm). Moreover the
wavelength ranges of the orders overlap (as usual for echelle
spectrographs).

In addition to slits with widths varying from
0$\stackrel{''}{\vspace*{-0.1ex}.}$4 to
5$\stackrel{''}{\vspace*{-0.1ex}.}$0 and with a common length of
11$\stackrel{''}{}$ X-Shooter also offers an Integral Field Unit of
1$\stackrel{''}{\vspace*{-0.1ex}.}$8$\times$4$\stackrel{''}{}$ that is
reformatted by mirrors into 3 slits of
0$\stackrel{''}{\vspace*{-0.1ex}.}$6$\times$4$\stackrel{''}{}$, which
are in turn aligned to form one long slit of
0$\stackrel{''}{\vspace*{-0.1ex}.}$6$\times$12$\stackrel{''}{}$.

The two optical arms (UVB/VIS) have atmospheric dispersion
compensators to reduce the effects of atmospheric refraction on the
observations.  In order to keep the targets at the same position along
the slit for each arm (despite the residual effects of flexure and
atmospheric refraction between the guiding wavelength and each arm's
central wavelength) each arm has a piezo controlled mirror. An arc
lamp exposure is taken before each observation and the observed
positions of a few carefully selected lines are compared to reference
positions. The offsets derived from these observations are then
compensated for by the piezo-controlled mirrors (Automatic Flexure
Control).

The lamps used for calibrations (wavelength calibration arc lamps and
flat field lamps) are mounted in the internal calibration unit. All
these components (slits, IFU, ADCs, mirror, calibration unit) as well
as the acquisition and guiding camera are part of the backbone, which
is mounted on the Cassegrain derotator of the telescope.

\section{QUALITY CONTROL}\label{sec:QC}

Quality Control (QC) in this article refers to the control of {\em
data quality}. This is part of the end-to-end data flow realized
within the ESO VLT, which is extremely important especially for
Service Mode (SM) observations.  SM observations are flexibly
scheduled to match prevailing ambient conditions. For a given
observing run, they might be scattered arbitrarily over a given
scheduling semester.  It is therefore of paramount importance that the
instruments used in such observations provide stable and reproducible
results. The SM paradigm assumes that if the instrument is healthy and
the PI defined constraints are fulfilled for the observation the data
will allow the astronomers to perform their scientific studies. The
instrument health is monitored mainly by QC using so-called Health
Check data, which are processed automatically by dedicated instrument
pipelines\cite{Modigliani,IZZO_2007} and which provide so-called QC parameters
like the measured central wavelength for a given setting, the detector
read-out noise, etc.  X-Shooter data are transmitted immediately after
acquisition at the VLT to the ESO headquarters at Garching. The
calibration part of the data flow is processed in an incremental
pattern, once per hour.  Quality information is fed back to Paranal
and reviewed to allow a very quick detection of possible problems.

Some Health Check data and parameters are of course very
instrument dependent, while others, like detector related parameters, 
may be common across several instruments. For spectrographs the
stability and reproducibility of the wavelength scale is one of the
most important aspects. Other parameters include instrumental
efficiency, stability of calibration lamps (to avoid underexposed data
as well as saturated ones), and detector parameters (e.g. detector
read-noise and presence of structure and/or patterns). 

\begin{figure}[h]
\begin{center}
\begin{tabular}{c}
\includegraphics[width=0.7\textwidth]{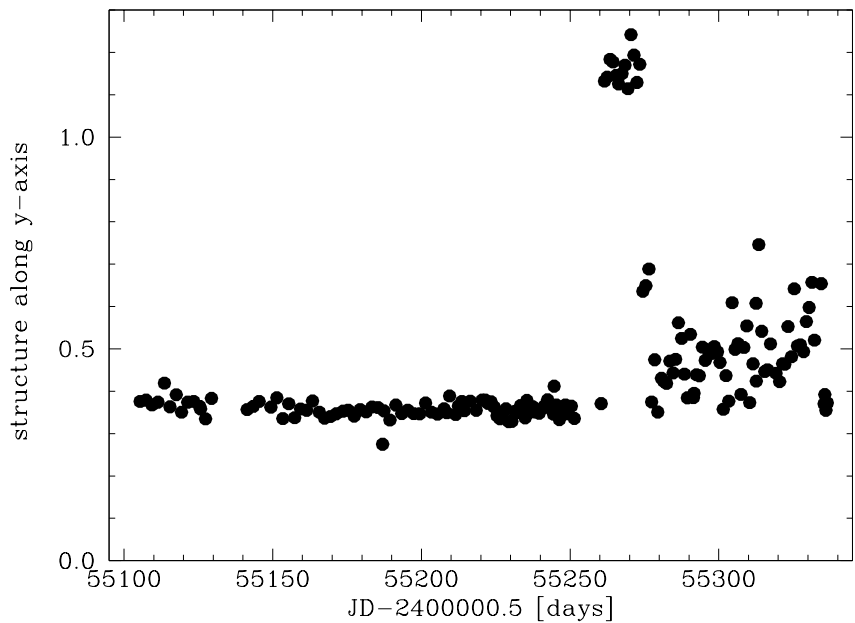}
\end{tabular}
\end{center}
\caption[]{The structure along the y-axis of the X-Shooter VIS CCD
  vs. time. The sudden strong increase marks the presence of
  an electronic interference pattern, which - albeit much reduced -
  persists until today.\label{Fig:bias_VIS} }
\end{figure}

Fig.~\ref{Fig:bias_VIS} shows a striking example of the usefulness of such
health check parameters. Here the structure along the y-axis is shown
for the fast, low gain readout of the X-Shooter VIS CCD. On March 6,
2010 (JD 2455262)
a strong interference pattern showed up for this readout mode,
which an be clearly seen in the plot as the data points above 1. While
the engineers managed to reduce the amplitude of this pattern to a
level where it is hard to see looking at the data, the elevated level of
the structure parameter still indicates its presence.


\section{PHYSICAL MODEL VS. PATTERN MATCHING}\label{sec:Model}
\begin{figure}[h]
\begin{center}
\begin{tabular}{c}
\includegraphics[width=0.9\textwidth]{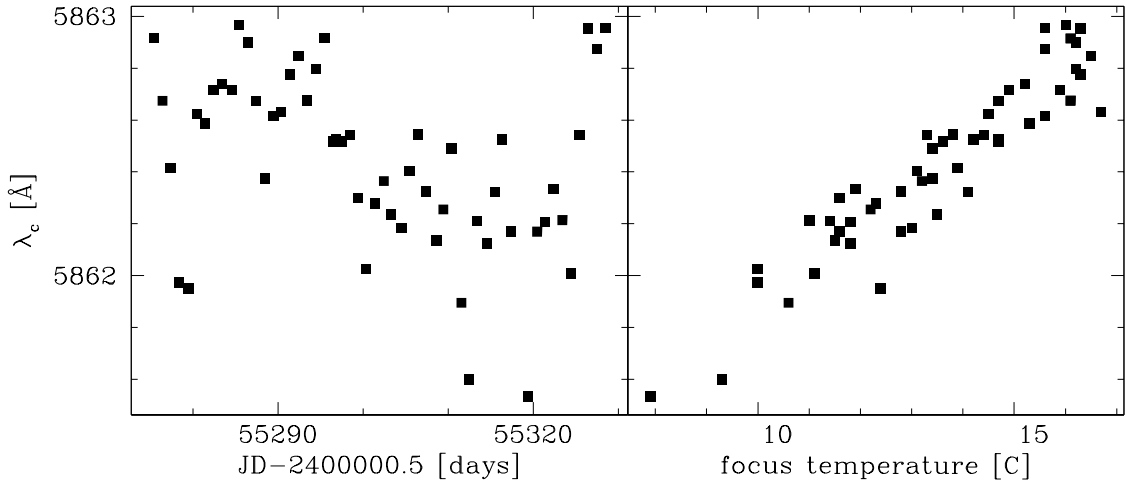}
\end{tabular}
\end{center}
\caption[]{The central wavelength of GRIS\_300V for the
  0.3$\stackrel{''}{}$ wide long-slit in FORS2 vs. time (left) and
  vs. focus temperature (right).\label{Fig:FORS2_LSS} }
\end{figure}

One of the main challenges for automatic data processing is the
correct identification of observed features like arc lines or
photometric standard stars. Here one can distinguish between two
principal approaches: pattern matching and physical model. Pattern
matching relies, as the name already suggests, on the detection of
patterns (e.g. groups of standard stars, groups of arc lines) to
correctly identify the observed features. This approach requires only
a rough idea of the instrumental characteristics (e.g. pixel scale,
dispersion, wavelength range) and allows to treat a large variety of
observations and instrumental configurations with the same
pipeline. It is therefore especially well suited to multi-mode
instruments (e.g. FORS2\cite{IZZO_2007}) and/or not very stable
instruments. In the case of FORS2 it has been shown that this approach
works extremely well and allows to process almost all of the
multi-object spectroscopic data\cite{FORS_2007}. As an impressive
example, tests showed that -- after adjusting the header keywords to
resemble FORS2 data -- the FORS2 pipeline processes multi-object
spectroscopy data from the Low-Resolution Spectrograph at the
Hobby-Eberly Telescope without problems. So the pattern matching
indeed provides very versatile routines. The disadvantage of this
approach is that the cause for a change in the observed data cannot
easily be identified. Fig.~\ref{Fig:FORS2_LSS} shows the central
wavelength of a specific long-slit setup for FORS2. Obviously the
central wavelength varies with time, but to identify the change in
temperature as the underlying cause was not trivial. And it still does
not tell whether the temperature affects the slit position or the
grism properties or both. The pattern matching also reaches its limits
in spectroscopy if there are large gaps between arc lines and/or the
edges of the nominal wavelength range are not covered by arc lines.

The physical model approach, on the other hand, relies on detailed
instrument knowledge and an appropriate implementation of that
knowledge in a pipeline. In this case the observed data are described
by a set of functions derived from the optical description of the
instrument (including the refractive indices of prisms etc.) and a
change in observed data may therefore more easily be traced back to a
change in instrument components. An additional advantage of the
physical model approach is that deficiencies in the calibration data
may be counterbalanced (e.g. a lack of arc lines in certain spectral
regions).  Also distortions can be much better described in such an
approach than with a standard polynomial fit, because they do not
depend in such a simple way on detector coordinates. Moreover the
polynomials often have boundary problems in the extreme areas (like
the edges of the field-of-view). The physical model is also extremely
well suited for the processing of echelle data, where the order
overlap requires that lines observed at rather different positions on the
detector are correctly assigned the same wavelength. While this often
causes problems in the classical polynomial fitting, where each order
is treated independently, the overall optimisation of the physical
model allows a comprehensive treatment of all lines simultaneously.

The physical model approach\cite{ROSA} was employed with great success
for the Space Telescope Imaging Spectrograph (STIS\cite{STIS1,STIS2})
onboard the Hubble Space Telescope. Independently ESO first adopted
this approach for the UV-Visual Echelle Spectrograph (UVES) in
2000\cite{Ballester}. Its pipeline uses a physical model to predict
and verify the spectral format and to allow robust and automatic
wavelength calibration for a virtually infinite number of instrument
settings (the central wavelengths of its independent blue and red arms
can be set arbitrarily within certain limits). At the time of the UVES
pipeline development and commissioning, the annealing functionality
was not as advanced, accurate, and reliable as it is now, thus in case
of large instrument shifts (e.g. caused by earthquakes), the
operational scenario requires a re-alignment of the instrument to a
reference position using reference spectral format check frames.

The physical model approach was then once more 
realized in the pipeline for the CRyogenic high-resolution InfraRed
Echelle Spectrograph (CRIRES\cite{CRIRES}), where it is especially
useful due to the limited coverage in wavelength per detector, which
can result in calibration data containing only one or even no arc
lines on a particular detector. X-Shooter is thus the third ESO
instrument for which a physical model is used in the automatic data
processing. In its pipeline\cite{Modigliani} -- if used in physical
model mode -- the model is used in most of the data reduction steps
(except master bias, dark and flat generation), including science data
processing.

\section{THE X-SHOOTER PHYSICAL MODEL}
The X-Shooter Physical Model (hereafter XPM) is a series of matrix
transformations, each representing an optical surface in the
spectrograph as specified in the Zemax optical design.  This enables
the physically based computation of the wavelengths associated with
each detector pixel in a given exposure based on a four-dimensional
vector containing the three spatial coordinates and the wavelength.

\begin{figure}[h]
\begin{center}
\begin{tabular}{c}
\includegraphics[width=0.6\textwidth]{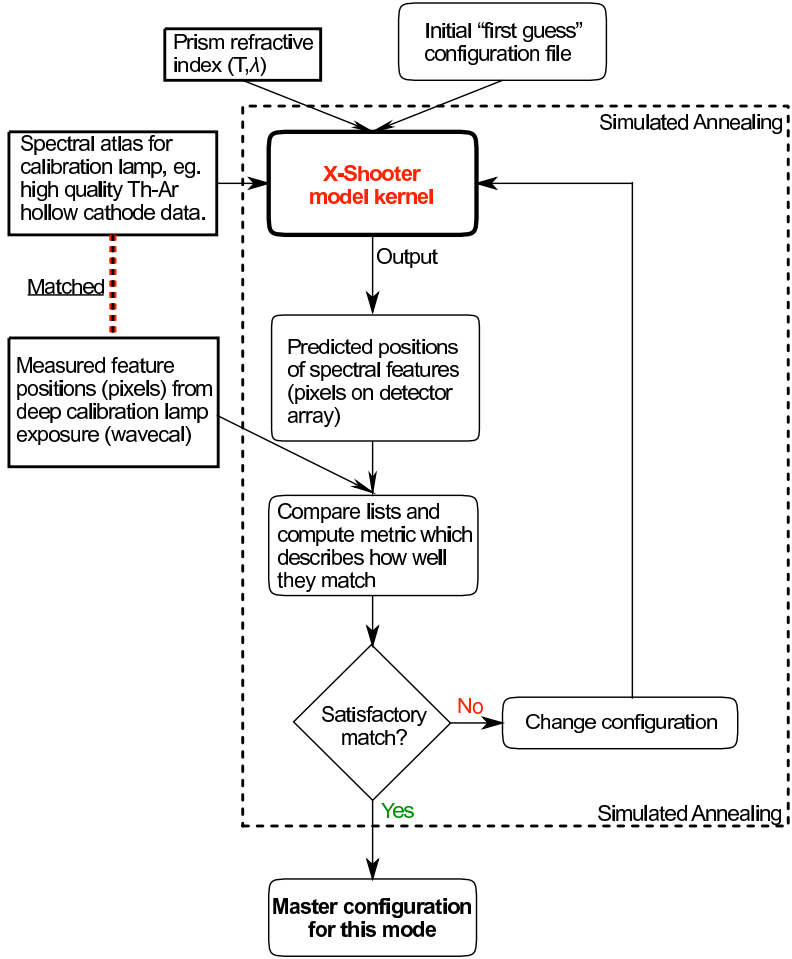}
\end{tabular}
\end{center}
\caption[]{The optimisation process of the physical model.\label{Fig:PhysMod} }
\end{figure} 

The central part of the XPM is a ray trace (referred to here as the
``model kernel'') that approximates the optical path of the three
X-Shooter spectrograph arms. Given a wavelength, echelle order and
a position along the slit, the model kernel will return the location at
which the detector is illuminated.

The model kernel is a simplified ray trace of the X-Shooter
spectrograph arms, based upon the optical design with the initial
values of many parameters taken directly from the X-Shooter Zemax
files. The simplification is necessary in order allow many iterations
of the model kernel both during the optimisation process or in some of
the higher level functions. Due to the similarity of the arms we are
able to use essentially the same model structure for all three arms.

The model kernel uses a parametrisation of the relative positions and
orientations of the principal dispersive optical components
(e.g. prisms) of the three X-Shooter arms.  We use exposures of
wavelength calibration sources (ThAr for UVB and VIS, ArKrNeXe penray
lamps for NIR), identify the locations on the detector of known
wavelengths and run an optimisation algorithm that finds the
combination of XPM parameter settings that is best able to place those
wavelengths at the observed positions. Figure~\ref{Fig:PhysMod}
provides a schematic overview of the optimisation process. As the
optimisation process is sensitive to false matches, we use customized
line catalogs that provide unblended lines at the X-Shooter spectral
resolution. Since the physical model needs fewer arc lines than the
polynomial approach the cleaning of the line lists proved to be
uncritical.

Once optimised in this way, the XPM and the accompanying parameter
configuration can be used to calibrate science exposures since the
wavelength associated with each pixel can be interpolated via
iterative executions of the model kernel. Moreover, as part of routine
operations, we continually re-optimise the XPM parameter
configuration. Obviously this gives us the most contemporaneous
calibration to apply to science data, and in addition it enables us to
monitor how the physical parameters of the instrument change.

\begin{figure}[!h]
\begin{center}
\begin{tabular}{c}
\includegraphics[width=0.9\textwidth]{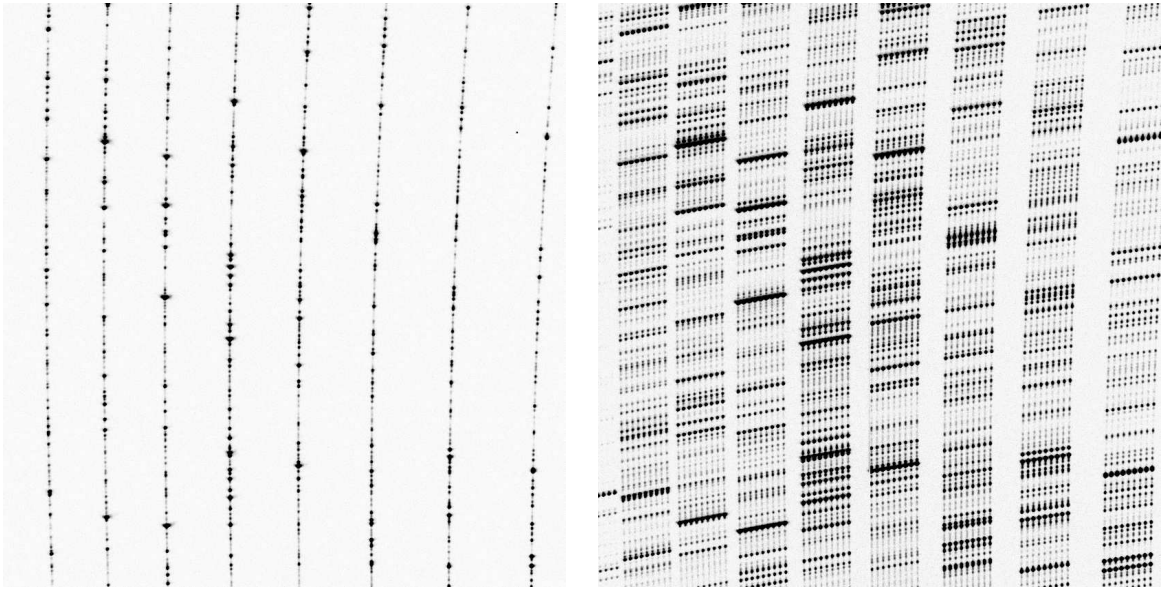}
\end{tabular}
\end{center}
\caption[]{UVB example of 1-pinhole (left) and 9-pinhole (right) data\label{Fig:distortion} }
\end{figure} 
The configuration files (one per arm) contain the following parameters
(number of parameters per arm in brackets). Fixed parameters are
marked by {\sl slanted font}:
\begin{itemize}
\item {\sl temperature of prism(s)} (read from header, 1 for UVB and VIS, 3 for NIR) 
\item orientation of entrance slit and detector plane along three
  axes (6)
\item orientation of prisms, for both entrance and exit surface, along
  three axes (6 for UVB and VIS, 18 for NIR)
\item orientation of grating along three axes (3) and grating constant (1)
\item location of the center of the pixel
  array along 2 axes and of the central pinhole/slit center  (2$+$2)
\item relative positions of the pinholes along the slit (8)
\item focal length of camera and collimator (2)
\item slit scale (1)
\item magnitude and wavelength zeropoint for chromatic aberration
  correction (2 for UVB and VIS)
\item additional tilt of primary NIR prism (1 for NIR)
\item {\sl pixel size} (1)
\item {\sl detector rotation} (1, only important for more than one
  detector per arm)
\item {\sl 2$^{\rm nd}$ order distortion coefficients}
\item {\sl refractive indices of the prisms} (1 each for UVB [Silica]
  and VIS [Schott SF6],  2 for NIR [Infrasil and 2$\times$ZnSe])
\end{itemize}
The fact that the prism temperatures are read from the headers allows
to adjust a configuration derived from daytime calibrations to
science data (and/or flexure compensation data) observed at night. 

\begin{figure}[h]
\begin{center}
\begin{tabular}{c}
\includegraphics[width=0.7\textwidth]{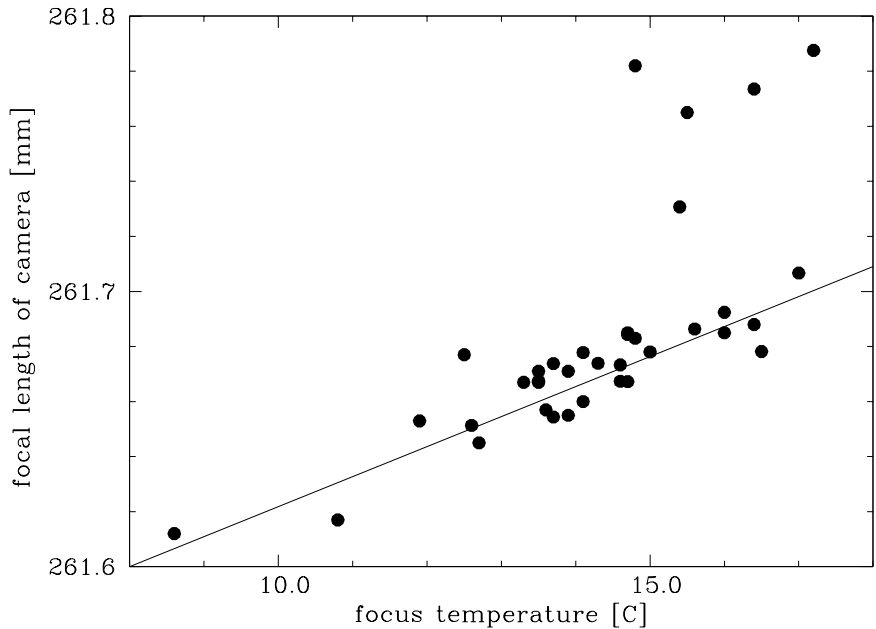}
\end{tabular}
\end{center}
\caption[]{The fitted focal length of the UVB
  spectrograph camera vs. the focus temperature used to adjust it. The
  line is {\em not} a fit to the data, but shows the relation used to
  mechanically adjust the focus (see text for
  details).\label{Fig:UVB_fdet} }
\end{figure} 

\section{APPLYING THE PHYSICAL MODEL}
We concentrate here on regular daytime calibration exposures obtained
with the telescope parked at Zenith (i.e. with the three
spectrographs in their null flexure position) and what they tell us
about physical parameters changing as a function of epoch and
environmental conditions. In a separate paper\cite{Bristow} we present a
similar analysis of physical parameters as a function of instrument
orientation.

Ultimately the goal of this work is firstly to gain some insight into
how the spectrographs change physically over time and secondly to be
able to predict changes in physical parameters as a function of
environmental conditions. This would allow the calibration to take
into account the fact that the environmental conditions at the time of
science exposures will often differ from those at the time of day time
calibrations.

\begin{figure}
\begin{center}
\begin{tabular}{c}
\includegraphics[width=0.9\textwidth]{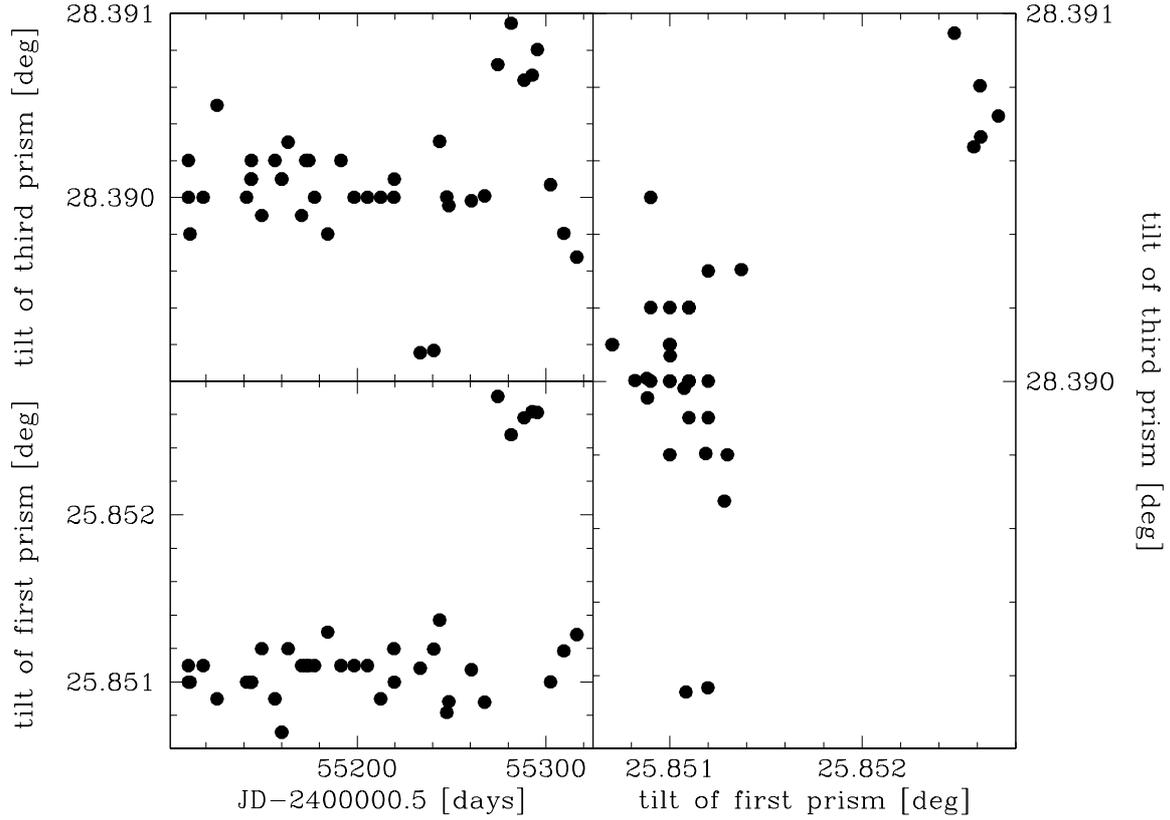}
\end{tabular}
\end{center}
\caption[]{The fitted orientation of the entrance
  surfaces of the first and third prism in the NIR
  arm vs. time (left) and vs. each other (right).\label{Fig:NIR_mup1_mup5} }
\end{figure} 

We use two types of X-Shooter calibrations to fine-tune the physical
model during normal operations (see Fig.~\ref{Fig:distortion} for
examples): 1-pinhole data (a.k.a. formatcheck) and 9-pinhole data
(a.k.a.  wave or 2dmap). About 40 such observations have been taken
since start of operations (October 1, 2009) for each arm and data
type. The 1-pinhole data are used to verify the overall format of the
data. Fitting these data the following parameters may be modified:
location of the central pinhole and of the central detector pixel,
grating constant, focal length of the camera, tip-tilt of all prisms
(both entrance and exit surfaces), orientation of the grating and of
the detector plane. If one fits the 9-pinhole data also the slit
scale, the orientation of the entrance slit and the focal length of
the collimator may be varied. 

In standard operations the parameter ranges are chosen so that they
are large enough to encompass any realistic physical changes over time
(in the absence of interventions or earthquake damage) but small
enough that the optimisation can be expected to converge without a
prohibitively high number of iterations. The ranges in use have been
established from an initial physically motivated estimate modified by
trial and error. An example of this is that we found that larger than
expected ranges were required for UVB 1-pinhole data, most likely due
to a not fully corrected temperature dependency. This process is
ongoing, and as we whittle down the number of open parameters we will
be less constrained by the optimisation time. To give an order of
magnitude, a typical parameter range is $\pm$0.1mm for the focal
distance of the VIS collimator (or $\approx$0.02\%) .

Figure~\ref{Fig:UVB_fdet} shows a good example of correlations found
with the physical model. The focus of the UVB camera is known to be
quite sensitive to temperature. It is therefore mechanically adjusted
automatically by 10.9$\mu$m per degree Celsius. The physical model
clearly finds exactly that correlation when fitting the data. The
nature of the 5 data points in the top right is unfortunately not yet
clear to us.
 
Due to the large number of parameter that are allowed to vary during
the optimisation we also sometimes encounter degenerate
parameters. Fig.~\ref{Fig:NIR_mup1_mup5} shows an example for such a
degeneracy with the orientation of the entrance surfaces of the first
and third prism in the NIR arm varying in step. In such cases the
algorithm finds a marginally better fit with varying two parameters,
which compensate each other to some degree. Such degeneracies are hard
to break with the current data set, which have for instance the same
flexure in all data. In a related paper\cite{Bristow} we describe
efforts to reduce the number of free parameters using flexure
compensation observations.

\section{SUMMARY}
The physical model used in the X-Shooter pipeline has proven to
provide an accurate description of the instrument, which allows to
process the vast majority of data without problems.  In addition it
provides us with information on the status of the various optical
instrument components and how that status changes with time and
temperature. Analysis of the currently available set of calibration
data has revealed some degeneracy amongst the physical model
parameters being optimised.  The next step will be to try to determine
a reduced set of parameters that is free of degeneracy. The
expectation is that this will reveal more robust correlations,
moreover the accumulation of data over time will improve the signal to
noise in the dataset.

\acknowledgments     
 
We thank P. Ballester and R. Hanuschik for their careful reading of
the manuscript and their helpful comments.


\bibliography{XSH} 
\bibliographystyle{spiebib} 

\end{document}